\documentclass{article}
\usepackage{arxiv}

\usepackage{hyperref}
\usepackage[linesnumbered,ruled,vlined]{algorithm2e}
\usepackage{array}
\usepackage[caption=false,font=normalsize,labelfont=sf,textfont=sf]{subfig}
\usepackage{textcomp}
\usepackage{stfloats}
\usepackage{url}
\usepackage{verbatim}
\usepackage{graphicx}
\usepackage{rotating}
\usepackage{balance}
\usepackage{multirow}
\usepackage{tabularx}
\usepackage{booktabs}
\usepackage{wrapfig}
\usepackage{caption}

\usepackage{xcolor}
\usepackage[normalem]{ulem}
\useunder{\uline}{\ul}{}

\newtheorem{Definition}{Definition}[section]
\newtheorem{theorem}{Theorem}[section]

\newtheorem{corollary}{Corollary}[section]

\newtheorem{proof}{Proof}[section]
\title{GraphC: Parameter-free Hierarchical Clustering of Signed Graph Networks}

\author{Muhieddine Shebaro, Lucas Rusnak, Martin Burtscher, and Jelena Te\v{s}i\'{c}}

\begin{document}

\maketitle
\begin{abstract} When applied to signed graphs, spectral clustering methods often struggle to capture natural groupings accurately. Recent research shows that these methods become less effective as the size of the signed network increases. To address this problem, we present a new, scalable, hierarchical graph-clustering algorithm called \emph{graphC} for identifying the best clusters in signed networks of various sizes. The \emph{graphC} clustering approach aims to keep the proportion of positive edges high within each cluster while maximizing the negative edges between clusters. One key feature of \emph{graphC} is that it does not require the number of clusters ($k$) to be predefined. We validate the performance of \emph{graphC} by comparing it to ten other signed-graph clustering algorithms on fourteen datasets. We show \emph{graphC} to be scalable to large signed graphs from Amazon, which include tens of millions of vertices and edges. On average, \emph{graphC} outperforms the second-best algorithm by 18.6\% in terms of maximizing the positive edge density within clusters and the negative edge density between clusters. This comparison excludes cases where the third-party algorithms fail to complete their tasks.
\end{abstract}
clustering, signed network, community detection, and structural balance.

%Motivation and Problem definition 
\section{Introduction}
\label{sec-problem}

Communities within a network are sets of vertices characterized by denser interconnections than with the broader network. These communities may exhibit distinct vertex sets or overlaps, where vertices participate in multiple communities. Detecting community structure is very important, as it gives us insights into relationships and dynamic mechanisms within complex networks spanning various domains, from biological to social networks. The inherent complexity of these networks frequently causes the resulting datasets to surpass the computational capacities of individual computers, making data partitioning for distributed processing a necessity.

The efficiency of such partitioning schemes in community-detection algorithms depends on various factors. The discovery of community structure is a challenging research topic, and numerous approaches have been proposed in the past that use local optimization~\cite{9146414}, statistical inference~\cite{POLLARD200299}, and machine learning~\cite{ALOMARY2006248}. While state-of-the-art community-discovery algorithms in unsigned graphs adeptly handle vast networks comprising millions of vertices and edges~\cite{edssjsED801E9B20231201}, the modeling of unsigned graphs falls short in capturing the complex relationships within the networks. Signed graphs capture positive and negative relationships well and offer more meaningful data representation~\cite{wang2022clust}. However, community discovery methods for signed network graphs struggle to recover communities in graphs featuring mere thousands of vertices and hundreds of thousands of edges~\cite{2022Cluster, wang2022clust}. The latest research underscores the challenges spectral methods pose in recovering the community structure in sparse networks, even with the incorporation of normalization techniques~\cite{2021cucuringu}. This finding aligns with our conclusions: there is a significant degradation in the performance of spectral methods for clustering large, sparse, signed networks derived from real-world data sources~\cite{2022Survey}. Balance theory is a pivotal concept in signed networks, explaining the evolution of attitudes within networks. Established by Heider~\cite{1958Abelson} and subsequently formalized mathematically by Harary, who introduced $k$-way balancing~\cite{Har2, Harary1968}, balance theory has found applications in predicting edge sentiment, content and product recommendations, and anomaly detection in various domains~\cite{derr2020link,garimella2021political,interian2022network,amelkin2019fighting}.

\subsection{Spectral Clustering of Real Signed Graphs} 

Donath and Hoffman~\cite{donath} introduced spectral clustering in 1973 to optimally partition the graph using the eigenvectors of the adjacency matrix. We outline the generic spectral clustering process in four steps: (1) computation of the Laplacian variant and identification of clusters ($k$); (2) derivation of $k$ eigenvectors corresponding to the $k$ smallest eigenvalues; (3) formation of the eigenvector matrix U to reduce dimensionality; and (4) application of $k$-means++ to cluster features. The SigNeT package, employed for spectral methods in signed networks, encompasses multiple Laplacian variants, all presumed to be positive semi-definite~\cite{aldogl2018}. The scalability of spectral clustering algorithms is limited in the context of real signed networks: (i) the substantial time required for the solver to formulate eigenvectors with their associated eigenvalues, and (ii) when dealing with large matrices, the approximations can become unstable and lead to errors due to something called spectral pollution or eigenvalue pollution~\cite{BOULTON20161}. A recent survey has provided a proof-of-concept and practical evaluation of these limitations in actual signed graphs~\cite{2022Survey}. The survey falls short of delineating the breaking points of signed Laplacians concerning algorithmic assumptions (e.g., small world and diameter) and characteristics specific to signed networks (e.g., density or sparsity)~\cite{2022Survey}.

A key limitation of state-of-the-art signed graph clustering algorithms is determining the optimal number of communities $k$ before running the algorithm. The elbow method is widely adopted, albeit subjective and not reliable~\cite{martinoElbow}. The eigenvector pollution, network sparsity, average degree, and overall structure can influence the choice of $k$ in real large graphs~\cite{2022Survey}. If the ground truth is not available, the metrics adopted might not be able to measure the quality of communities discovered by the methods~\cite{2022Survey}. Thus, parameter-free techniques are preferable. 

\subsection{Research Contributions}
\label{sec-obj}

This paper introduces a novel \emph{graph clustering} algorithm called \emph{graphC}. The main contributions are: 

\noindent $\bullet$  We prove that Harary cuts from a \emph{balanced} signed network are equivalent to an eigenvector with a zero eigenvalue of that balanced network.

\noindent $\bullet$ The \emph{graphC} algorithm effectively implements this new duality finding by directly performing Harary cuts on the balanced signed graph.

\noindent $\bullet$ The \emph{graphC} algorithm is $k$-independent: it does not require a predefined number of clusters to produce a high-quality clustering. The parameters used in the algorithm are dependent on computing resources. 

\noindent $\bullet$ The \emph{graphC} algorithm identifies optimal clusters without requiring a spectral decomposition solver. The algorithm is eigenvalue pollution-free for large sparse matrices. 

\noindent $\bullet$ The \emph{graphC} addresses the positive/negative edge imbalance that naturally occurs in most signed networks by equally considering the contribution of positive and negative edges to optimize the clustering.

\section{Related Work}
\label{sec-related}

Existing approaches for community detection in signed networks form a diverse landscape. Xhiang et al.~propose an algorithm for clustering signed graphs utilizing the balanced normalized cut, demonstrating commendable clustering results yielding an efficient approach~\cite{2012chiang}. The \emph{Signed Positive over Negative Generalized Eigenproblem} spectral-based approach necessitates the specification of $k$ since it formulates clustering as a generalized eigenvalue problem, enhancing a well-defined objective function~\cite{pmlr-v89-cucuringu19a}. Relying on the signed Hessian for clustering also requires a predefined $k$. It emphasizes the advantages of non-backtracking operators with the computational and memory efficiencies inherent in real symmetric matrices~\cite{Saade2014SpectralCO}.

He et al.~\cite{he2022sssnet} use modified social network analysis and triangle balancing heuristics (``the friend of my friend is my friend'') to address the issue of cluster discovery based on a modified version of Heider's balance theory. First, the authors employ a signed mixed-path aggregation (SIMPA) method to construct the vertex embedding. Consequently, this vertex embedding produces probabilities for cluster assignments. The clustering process involves training with a weighted combination of supervised and unsupervised loss functions, with the unsupervised component being a probabilistic balanced normalized cut. The algorithm uses SPONGE (SPO) labels for some signed graphs to execute the clustering, building a dependency on other signed graph clustering algorithms. Mercado et al.~\cite{means} propose to combine the positive Laplacians (part of the Laplacian matrix with positive edge weights) with negative Laplacian (part of the Laplacian matrix with negative edge weights) using matrix power means. The result is a modified version of the Laplacian matrix that considers both positive and negative relationships in the graph. Next, homophily is the tendency of a vertex to connect to similar ones, and negative connections introduce noise as those connections are to dissimilar vertices. The solutions to design the graph neural networks (GNN) for signed network clustering utilizing the balance theory showed initial promise~\cite{gnnasine, bridge}. However, the issue of misrepresenting the negative edges in the resulting embedding proved to be a roadblock to further development of community discovery and recommender systems in this direction~\cite{convsigned,siren}. The \emph{neutrosophic} theory, introduced in~\cite{overlapping}, proposes a novel signed graph convolutional network (SGCN) to encapsulate better the architecture of the signed network in the compact space and overcome the embedding issues. The clustering in the neutrosophic feature space identifies overlapping communities in signed networks~\cite{overlapping}.

\begin{figure*}[!ht]
  \centering
  \includegraphics[scale=0.45]{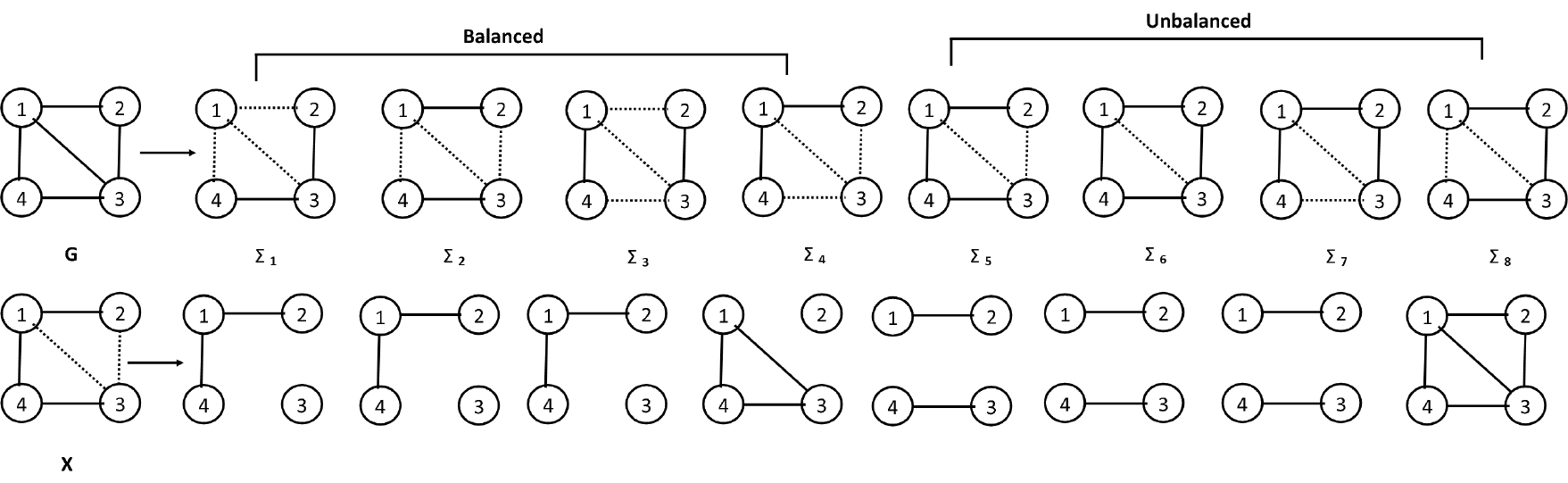}
  \caption{\textbf{Dotted} lines are negative edges whereas \textbf{sold} lines are positive edges. Top: A graph $G$, a balanced graph $\Sigma_1$ obtained by switching (changing the sign of the edges originating in) $v_1$, a balanced graph $\Sigma_2$ obtained by switching $v_1$ and $v_2$, a balanced graph $\Sigma_3$ by switching $v_2$ and $v_3$, and a balanced graph $\Sigma_4$ by switching $v_4$. $\Sigma_5$, $\Sigma_6$, $\Sigma_7$ and $\Sigma_8$ are examples of unbalanced states. Bottom: Harary Cuts of nearest stable states~\cite{2021Cloud} of $\Sigma_5$.}
  \label{fig-SpecClustGraphs}
\end{figure*}

\section{Balanced-Spectral Duality in Signed Networks}
\label{sec-Duality}

In this section, we present the terminology and properties of signed graphs and introduce spectral analysis and the duality theory of near-balanced states. Let $\Sigma_i$ be a signed graph with the unsigned topology $G$, as illustrated in Figure~\ref{fig-SpecClustGraphs}. Let $\Sigma_{ij}$ be the collection of vertex sets discovered by the clustering algorithm $j$. The \emph{graphC} algorithm minimizes the fraction of positive edges \emph{between} communities and the fraction of negative edges \emph{within} communities for algorithm $j$ applied to the signed graph $\Sigma_i$. We quantify the positive edges outside communities as $pos_{out}(\Sigma_{ij})$  and the negative edges within communities as $neg_{in}(\Sigma_{ij})$ for algorithm $j$ operating on the signed graph $\Sigma_i$.

\subsection{Fundamental Cycle Basis}
\begin{Definition}
Graph $\Sigma_i$ is a \textbf{subgraph} of a graph $G$ if \textbf{all} edges and vertices of $\Sigma_i$ are contained in $G$. \label{def:Subgraph}
\end{Definition}

\begin{Definition}
A \textbf{path} is a sequence of $m$ edges that connect a sequence of $n$ vertices in a graph. A \textbf{connected graph} has a path between every pair of vertices. A \textbf{cycle} is a path that begins and ends at the same vertex. A \textbf{cycle basis} is a set of simple cycles that forms a basis of the cycle space (see below). \label{def:CycleBasis}
\end{Definition}
\begin{Definition}
Let $T$ be a spanning tree of the underlying graph $G$, and let $e$ be an edge in $G$ between vertices $x$ and $y$ that is \emph{NOT} in the spanning tree $T$. Since the spanning tree spans all vertices, the unique path in $T$ between vertices $x$ and $y$ does not include $m$. A \textbf{fundamental cycle} in $G$ is the cycle that is created when adding edge $e$ to this path in $T$. \label{def:FundamentalCycle}
\end{Definition}

\begin{corollary}
For a given connected graph $G$, we create the fundamental cycle basis when we select all cycles formed by combining a path in the tree and a single edge outside the tree. For a graph $G$ with $n$ vertices and $m$ edges, there are exactly $m-n+1$ fundamental cycles because the spanning tree includes $n-1$ edges.
\end{corollary}

\subsection{Balanced Signed Graphs}
\label{ssec-Balance}
\begin{Definition}
The \textbf{signed graph} $\Sigma=(G, m)$ consists of the underlying unsigned graph $G$ and an edge signing function $m \rightarrow \{+1,-1\}$. The edge $m$ can be positive $m^+$ or negative $m^-$. The \textbf{sign} of a sub-graph is the \emph{product} of the signs of its edges. A \textbf{balanced signed graph} is a signed graph where every cycle is positive. The \textbf{frustration} of a signed graph is defined as the minimum number of candidate edges whose signs need to be switched for the graph to reach a balanced state. Figure~\ref{fig-SpecClustGraphs} (top, balanced states) provides an example.
\label{def:SignedGraph}
\end{Definition}

\begin{theorem}[\cite{Har2}] If a signed subgraph $\Sigma$ is balanced, the following are equivalent:
\begin{enumerate}
 \setlength{\leftmargin}{0pt}
  \item $\Sigma$ is balanced (all circles are positive).
  \item For every vertex pair $(v_i, v_j)$ in $\Sigma$, all $(v_i, v_j)$-paths have the same sign.
  \item $Fr(\Sigma) = 0$ (the frustration of $\Sigma$ is zero).
  \item The vertices in the network can be grouped into two sets $U$ and $W$ where the edges between the vertices within the sets $U$ and $W$ are positive and the vertices between the sets are negative. $(U,W)$ bi-set is called Harary-bipartition.
\end{enumerate}
\label{t:HararyCut}
\end{theorem}

The formation of the Harary cut starts with deleting the negative edges in each balanced state in the frustration cloud of the signed graph, as illustrated in Figure~\ref{fig-SpecClustGraphs} (bottom). This figure's trivial unbalanced signed graph consisting of 4 vertices and five edges yields eight nearest balanced states. Note that a Harary cut can yield multiple connected components, and the input graph might contain many initially connected components where Harary cuts and balancing take place in every element. In prior work, we have characterized signed graphs through the frustration cloud~\cite{2021Cloud}, a collection of nearest balanced states, and then introduced an improved and parallelizable way to efficiently discover the nearest balanced states in large signed graphs~\cite{2021Alabandi}.

\subsection{Spectral Clustering and Balanced States}

A signed graph is \emph{balanced} if the graph has no cycles with an odd number of negative edges. The \emph{switch} operation flips the sign of all edges connected to a specific vertex as shown in Figure~\ref{fig-SpecClustGraphs} when switching $v_1$ in $G$ to obtain $\Sigma_1$. Switching equivalence in this context means that two balanced signed graphs can be transformed into each other through a series of switch operations, as shown in the following example. The balanced signed graphs of the same underlying graph $G$ in Figure~\ref{fig-SpecClustGraphs} (Top) are switching equivalent~\cite{Marsden2013EIGENVALUESOT}. For example, $\Sigma_2$ and $\Sigma_3$ in Figure~\ref{fig-SpecClustGraphs} (Top) are switching equivalent because we can obtain $\Sigma_3$ by switching $v_1$ and $v_3$ in $\Sigma_2$. The key point is that switching operations preserve the balanced nature of the graph as they do not alter the overall balance or imbalance of the graph~\cite{2021Cloud}. We get the same eigenvalues, but the eigenvectors only differ by a multiple of $-1$ for each vertex switched.

\begin{table}[!ht]
\caption{Balanced signed graphs $\Sigma_0$, $\Sigma_1$, and $\Sigma_2$ are isospectral: they have identical eigenvalues, but their eigenvectors differ. $\Sigma_5$ is an unbalanced graph, so its eigenvalues and eigenvectors differ.}
\label{tab-isospectral} %\footnotesize
\setlength\tabcolsep{1pt}
\centering
\begin{tabular}{l|r|r|r|r|r|r|r|r|r|r|r|r|r|r|r|r}
& \multicolumn{12}{c|}{Balanced Graphs} & \multicolumn{4}{c}{Unbalanced Graph} \\ \hline
Graph & \multicolumn{4}{c|}{$\Sigma_0$} & \multicolumn{4}{c|}{$\Sigma_1$} & \multicolumn{4}{c|}{$\Sigma_2$} & \multicolumn{4}{c}{$\Sigma_5$} \\ \hline
Eigen & $0$ & $2$ & $4$ & $4$ & & & & & $0$ & $2$ & $4$ & $4$ & & & & \\
values & & & & & $0$ & $2$ & $4$ & $4$ & & & & & $2-\sqrt{2}$ & $3-\sqrt{3}$ & $2+\sqrt{2}$ & $3+\sqrt{3}$ \\ \hline
& 1 & 0 & 1 & 0 & 1 & 0 & 1 & 0 & 1 & 1 & 1 & 1 & 1& 1& -1 & 1\\ 
Eigen & 1 & 1 & 0 & 1 & -1 & 1 & 0 & 1 & 1 & 1 & 0 & 1 & $\sqrt{2}$ & 0& $\sqrt{2}$& 0 \\ 
vectors& 1 & 0 & -1 & -2 & -1 & 0 & 1 & -2 & -1 & 0 & 1 & 2 & -1 & 1& 1 &1\\
 & 1 & -1 & 0 & 1 & -1 & -1 & 0 & 1 & -1 & 1 & 0 & -1 & 0 & $\frac{1+\sqrt{3}}{2}$ & 0 & $\frac{1-\sqrt{3}}{2}$\\
\end{tabular}
\end{table} Note that the spectral clustering yields as many zero eigenvalues as there are clusters in the data~\cite{Marsden2013EIGENVALUESOT}. The zero eigenvalues have an eigenspace with a basis of vectors that only have zeros and ones as elements, and these vectors indicate which vertices belong to each cluster~\cite{Marsden2013EIGENVALUESOT}. In Table~\ref{tab-isospectral}, for the connected graph G of 4 vertices with all positive edges, the eigenvector with the zero eigenvalue consists entirely of ones. All balanced signed graphs are equivalent to all positive signed graphs with the same topology. Since the Laplacian matrices of balanced signed graphs of the same underlying graph are isospectral. Table~\ref{tab-isospectral} exemplifies this using the graph G from Figure~\ref{fig-SpecClustGraphs} (Top, balanced states) whereas the $\Sigma_5$ column in Table~\ref{tab-isospectral} shows the spectral clustering of an unbalanced state.

Thus, when performing spectral analysis on such graphs, the eigenvalues remain the same, as illustrated in Table~\ref{tab-isospectral}. The corresponding eigenvectors only differ by a multiple of -1 for each switched vertex, as illustrated in Table~\ref{tab-isospectral} for the eigenvalues and eigenvectors of $G$, $\Sigma_1$, and $\Sigma_2$ (from Figure~\ref{fig-SpecClustGraphs}). This property makes the study of spectral characteristics more manageable, allowing for a direct comparison between balanced signed graphs while preserving their essential structural features. The multiplicity of the $0$ eigenvalue counts the number of connected components of the underlying graph when balanced. To see this fact, let $W \subseteq V$ be the set of vertices switched from the original, all-positive graph $G$, that produces another balanced graph $\Sigma$. Let $\mathbf{I}_{W}$ be the $V \times V$ diagonal matrix whose $(v,v)$-entry is $-1$ if $v \in W$ and $+1$ otherwise. Thus, the Laplacian for the balanced signed graph $\Sigma$ is $\mathbf{L}_{\Sigma} = \mathbf{I}_{W}\mathbf{L}_{G}\mathbf{I}_{W}$.

We also present a simple new proof of the isospectrality of balanced states, highlighting the signed cycle structure and further indicating the need for non-spectral methods for sentiment analysis. 

\begin{theorem}
  The Laplacian matrices of balanced signed graphs of the same underlying graph are isospectral. 
\end{theorem}
\begin{proof}
  First, observe that switching does not change the cycle signs. This is because every vertex in a given circle has a degree equal to 2 in that circle subgraph. Switching reverses exactly two edge signs, which produces no change in the cycle sign. Next, we use the characteristic polynomial formulation for the Laplacian from~\cite{OHSachs}. In there, it was shown that the signs of the cycle subgraphs entirely determine the characteristic polynomial. Since each balanced signed graph of an underlying graph has all cycles positive, they must necessarily have the same characteristic polynomial and, hence, the same eigenvalues. 
\end{proof}

\begin{theorem}
  For a connected balanced signed graph $\Sigma$, the entries of the $0$-eigenvector are $+1$ or $-1$. Partitioning the vertices along the entry values of the $0$-eigenvector corresponds to the Harary cut of the balanced signed graph.
\end{theorem}
\begin{proof}
  Since $\Sigma$ is balanced, it is isospectral to the underlying graph $G$. Since $G$ is connected, the dimension of the $0$-eigenspace is $1$, and the vector $\mathbf{1}$ consisting of all $ 1$'s is a basis for that eigenspace of $G$. Switching a vertex $v$ negates both row and column $v$ in $\mathbf{L}_{G}$ to produce $\mathbf{L}_{\Sigma}$, thus negating the $v$-entry of the basis vector $\mathbf{1}$. 
\end{proof} 

Note that since an unbalanced signed graph contains negative cycles, the proof of the previous theorem ensures that these cycles must have different eigenvalues. Therefore, if a signed graph is unbalanced, $0$ is not an eigenvalue, and the structure and sentiment of the graph effectively merge into its geometry, as shown in Table~\ref{tab-isospectral} for $\Sigma_5$. In contrast, balanced graphs share the same eigenvalues and allow multiple optimal ways to divide the graph into two components. Spectral methods rely on the geometry of the eigenvectors rather than the sentiment since the sentiment is embedded in the signs of the vectors and not in the proximity of the data (such as in a $k$-means approach). Thus, we conclude that both the Harary cut and spectral methods involve bisecting the graph. We can create a balanced Laplacian matrix once a signed network is balanced. The spectral decomposition of this matrix reveals the eigenvectors and their eigenvalues. If the graph is balanced and connected, there will be a $1$-dimensional kernel, and the corresponding $0$ eigenvalue will provide the Harary cut.

\section{Measuring Signed Graph Clustering Efficacy}
\label{sec-measure}
\begin{figure}[!ht]
  \centering
  \includegraphics[width=3.1in]{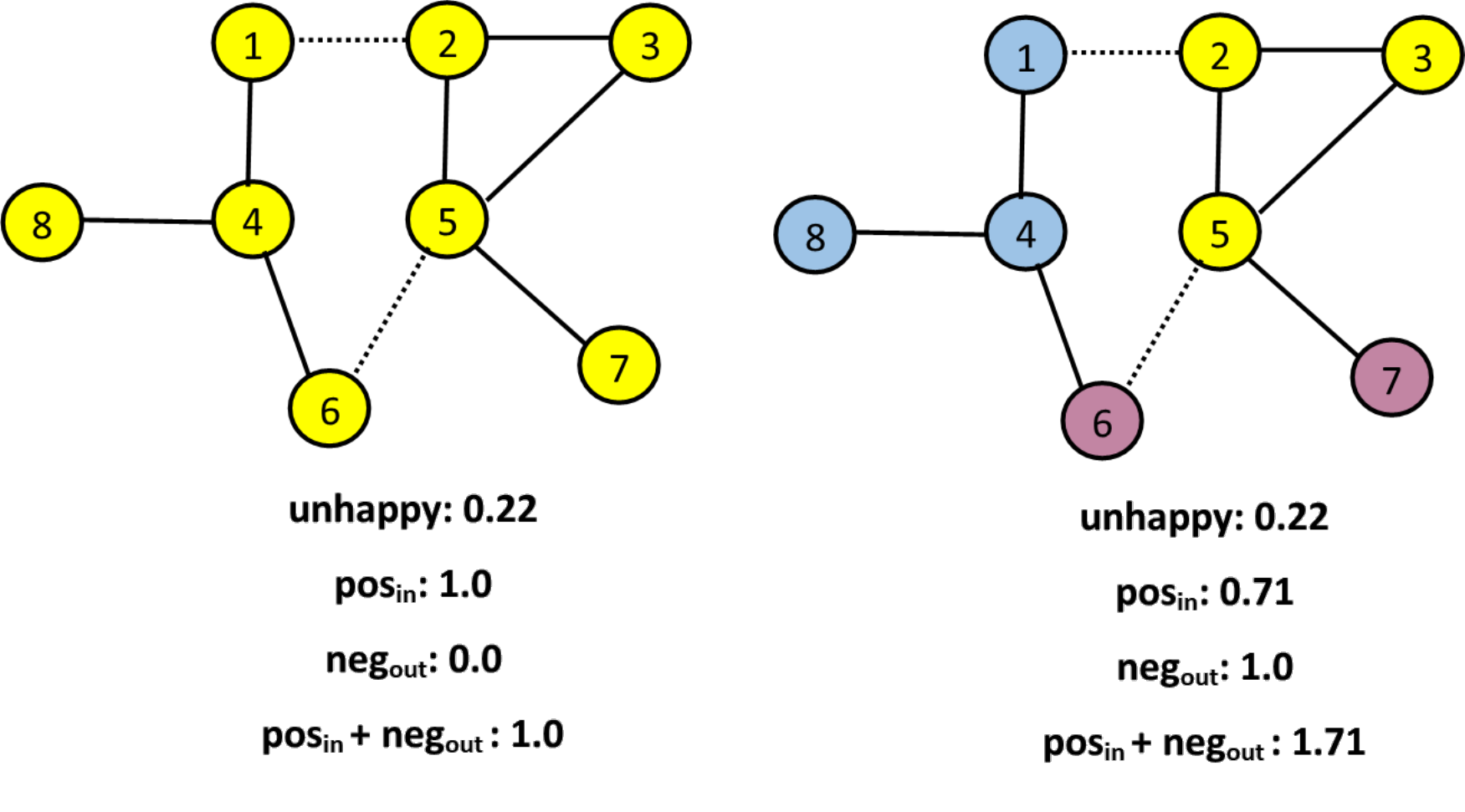}
  \caption{An illustration emphasizing the advantage of measuring the quality of a clustering assignment in a signed graph using the summation of the fraction of positive edges within communities and the fraction of negative edges between communities. \textbf{Dotted} lines are negative edges whereas the \textbf{sold} lines are positive edges.}
  \label{fig-example_unhappy}
\end{figure} Consider two clustering assignments for the network illustrated in Figure~\ref{fig-example_unhappy}. Without ground truth and training data for the community assignment in signed networks, the state-of-the-art uses the \emph{unhappy ratio} to determine the quality of the clustering~\cite{he2022sssnet}. The \emph{unhappy ratio} $U_i$ is defined in Eq.~\ref{eq-un} for a set of communities produced by method $j$ in the signed graph $\Sigma_i$: \begin{equation} \label{eq-un}
U_{ij}=\frac{pos_{between}+neg_{within}}{pos_{between}+ pos_{within}+neg_{between}+neg_{within}}
\end{equation}
\noindent The $pos_{between}$, $pos_{within}$, $neg_{within}$, and $neg_{between}$ are the number of positive edges between communities, the number of positive edges within communities, the number of negative edges within communities, and negative edges between communities, respectively. 

The \emph{unhappy} ratio for both clustering approaches in Figure~\ref{fig-example_unhappy} is $0.22$. The clustering assignment on the right seems more effective as it avoids disregarding \emph{all} negative edges, successfully segregates two communities (1,4,8,6 and 2,3,5,7), and accurately classifies \emph{most} positive edges. The \emph{unhappy} ratio is unable to make these distinctions. It fails to capture the quality of the clustering of real signed networks because the \emph{unhappy} ratio favors positive edges and trivial clustering cases in networks where the typical number of negative edges is 5-10 times smaller than the number of positive edges~\cite{2022Survey}. 

Consider another signed graph with 1000 edges, where 900 are positive, and 100 are negative, and two cluster assignments. In the first clustering, there are 400 positive edges between clusters (\( pos_{between} \)), 500 positive edges within clusters (\( pos_{within} \)), 50 negative edges within clusters (\( neg_{within} \)), and 50 negative edges between clusters (\( neg_{between} \)), yielding an unhappy ratio (\( U_1 \)) of 0.45. The second clustering has zero positive edges between clusters, 900 positive edges within clusters, 100 negative edges within clusters, and zero negative edges between clusters, resulting in an unhappy ratio (\( U_2 \)) 0.1. In this example, the distribution of positive and negative edges is imbalanced. The first clustering can separate some negative edges into different clusters while preserving positive edges within the same clusters. The second clustering is trivial because it places all vertices in a single cluster. Thus, the \emph{unhappy ratio} as defined favors a trivial class when the number of positive edges is much higher than the number of negative edges and vice versa. In the edge sign imbalance case, the \emph{unhappy ratio} is a low value for the clustering approach that puts all vertices into one community. As a remedy, we change the \emph{unhappy ratio} to the \emph{unhappy score}, which accounts for both positive and negative edges as follows:
\begin{equation}
US_{ij} = \frac{pos_{between}}{pos_{between}+pos_{within}}+\frac{neg_{within}}{neg_{within}+neg_{between}}
\end{equation} Intuitively, measuring the sum of the fraction of positive edges within communities (\( pos_{in} \)) and the fraction of negative edges between communities (\( neg_{out} \)) provides a better assessment of the clustering quality as shown in Figure~\ref{fig-example_unhappy}~\cite{2022Survey}. The unhappy ratio for both clustering assignments is the same (0.22), even though the left assignment ignores all negative edges, while the right assignment considers both positive and negative edges. In Figure~\ref{fig-example_unhappy}, the left assignment has an updated $pos_{within}+neg_{between}$ of 1.0. The right assignment has an updated $pos_{within}+neg_{between}$ of 1.71, and the scores reflect better the true clustering quality. 

The updated \emph{unhappy score} is the new \emph{graphC} loss measure for a signed graph $\Sigma_i$ with an unbalanced edge ratio and clustering algorithm $j$. We rewrite it as a loss function since our objective in clustering is to minimize $\mathcal{L}(\Sigma_{ij})$ by splitting vertices into communities. The $pos_{out}$ and $neg_{in}$ are defined as
\begin{equation}
  pos_{out}=\frac{pos_{between}}{pos_{between}+pos_{within}}~~~~neg_{in}=\frac{neg_{within}}{neg_{within}+neg_{between}}
  \label{eq-PosOutNegIn}
\end{equation}
Thus, we propose to minimize the fraction of violating negative edges $neg_{in}(\Sigma_{ij})$ as well as the fraction of violating positive edges $pos_{out}(\Sigma_{ij})$ simultaneously. In this context, we define the violating edges ($\mathcal{V}$) as the total number of positive edges between clusters and negative edges inside them.
\begin{equation}
\mathcal{V}_{ij} = \mathcal{L}_{\Sigma_{ij}}=pos_{out}+neg_{in}
\label{eq-UnPlus}
\end{equation}

Symmetrically, we redefine $pos_{in}$ (Eq.~\ref{eq-PosIn}) and $neg_{out}$ (Eq.~\ref{eq-NegOut}) measures in terms of fractions to account for the positive and negative edge imbalance. 
\begin{equation}
 pos_{in}= \frac{pos_{within}}{pos_{between}+pos_{within}}
  \label{eq-PosIn} 
\end{equation}
\begin{equation}
neg_{out}= \frac{neg_{between}}{neg_{within}+neg_{between}}
  \label{eq-NegOut} 
\end{equation}

Next, we generalize the loss function to accommodate for the graph characteristics:
\begin{equation}
\label{eq-Loss}
\mathcal{L}_{\Sigma_{ij}}(\alpha,\beta)=\beta(\alpha{pos}_{out}+(1-\alpha){neg}_{in})+(1-\beta)*\frac{|V_{iso}|}{|V|} 
\end{equation}

\noindent The $\alpha$ controls the importance between ${pos}_{out}$ and ${neg}_{in}$. Setting $\alpha =$ 0.5 gives equal importance to both. The $\beta$ parameter controls the fraction of isolated vertices $|V_{iso}|$ produced in the process; $|V|$ is the total number of vertices in the graph. If $\beta = 0$, then the algorithm will completely ignore minimizing the ${pos}_{out}$ and ${neg}_{in}$ and will find the cuts that yield the least amount of isolated vertices possible. Note that $\mathcal{L}_{\Sigma_{ij}}$ in Eq.~\ref{eq-UnPlus} is equal to $\mathcal{L}_{\Sigma_{ij}}(0.5,1)$ in Eq.~\ref{eq-Loss}. 

\begin{figure*}[!t]
  \centering
  \includegraphics[width=\textwidth]{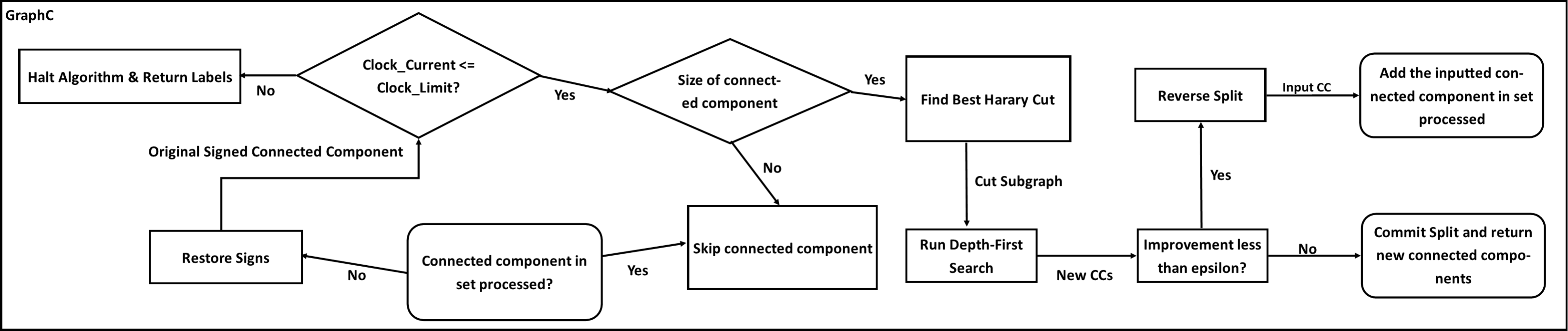}
  \caption{The \emph{graphC} pipeline: the starting state is ``Connected component in set processed?'' block.}
  \label{fig-pipeline}
\end{figure*}

\section{The \emph{graphC} Methodology}
\label{sec-GraphC}

Section~\ref{sec-Duality} showed that the Laplacian matrices of balanced signed graphs with the same underlying graph are isospectral and that spectral clustering reconstructs a Harary cut. The proposed \emph{graphC} algorithm searches for the optimal Harary cut using the quality measure defined in Equation~\ref{eq-Loss}, along with a set of stopping criteria to automatically halt the algorithm and return the final clustering labels for each vertex. Figure~\ref{fig-pipeline} outlines the flow of the \emph{graphC} algorithm. It starts by labeling all vertices within the same connected component with the same label, creating an initial clustering assignment set \(\mathcal{C}\), as described in Algorithm~\ref{alg-label}. Isolated vertices receive unique labels, and each isolated vertex is assigned a different label. Specifically, the algorithm first checks whether the connected component is in a set called \(processed\). If not, it restores the original signs using the signed network before continuing. Next, the algorithm checks a time limit and ensures that the size of the connected component is greater than the threshold \(\Gamma\). Once the algorithm identifies the best Harary cut based on the loss function in Equation~\ref{eq-Loss}, it evaluates whether the Harary split improves the overall clustering quality, as measured by Equation~\ref{eq-overall_loss}. Finally, if the improvement exceeds a threshold \(\epsilon\), the algorithm commits the split; otherwise, it terminates.

\begin{minipage}{0.48\textwidth}
\begin{algorithm}[H]
\SetAlFnt{\small\sf}
\SetAlgoLined
\caption{Split and Label Components}\label{alg-label}
\KwData{$\Sigma_{ij}$, label\_counter}
\KwResult{Clustering label assignment set $\mathcal{C} = \{C_v\}, v \in V$}
label\_counter=$0$\;
\For{$CC$, $CC \in \Sigma_{ij}$}{
  \For{ $v$, $v \in CC$}{
    ${C_v}$=label\_counter\;
  }
  label\_counter++\;
}
\end{algorithm}
\end{minipage}
\hfill
\begin{minipage}{0.5\textwidth}
\begin{algorithm}[H]
\SetAlFnt{\small}
\SetAlgoLined
\caption{Best Harary Cut}
\label{alg-bestsplit}
\KwData{$\Sigma_{ij}$, Set R, label\_selected, $\mathcal{C}$, $\alpha$, $\beta$, $I$}
\KwResult{$G^{FC}$}
\For{every vertex $v$ $in$ $\Sigma_{ij}$}{
\If{$C_v != $ label\_selected}{
 R = R + $v$ \;
}
}
$ G^{F} = G \setminus R $\; 
$\Sigma=$ GraphBplus($G^F,I$) \; 
$G^{FC} = \Sigma | \{min \mathcal{L}_{\alpha,\beta,G^F}\}$\;
\end{algorithm}

\end{minipage}

\textbf{Selecting the Best Harary Cut:}
We use the graph plus algorithm to create stable states and perform Harary cuts. To find the best Harary cut, we pick the one that results in the slightest loss, as defined in Eq.~\ref{eq-Loss}. After selecting the Harary cut, we employ Depth-First Search (DFS) to create the connected components. DFS starts at the root vertex and explores each branch as far as possible before backtracking. The process continues until \emph{graphC} has visited all vertices connected to the root vertex. The time it takes for DFS to run is proportional to the number of vertices ($|V|$) plus the number of edges ($|E|$). Algorithm~\ref{alg-bestsplit} demonstrates the process of selecting the best Harary cut.

\textbf{Stopping Criteria:} After finding the best Harary cut of a particular connected component, the algorithm computes the overall quality of the new clusters of the entire signed network, which is similar to Equation~\ref{eq-Loss} and works as follows: \begin{equation}
\label{eq-overall_loss}
\mathcal{L}_{\Sigma_{ij}}^{t}=({pos}_{out}+{neg}_{in})
\end{equation}

The $epsilon$ parameter is a defined stopping condition for the graph clustering. If the new clusters result in an improvement of less than $epsilon$, the algorithm undoes the last split of the connected component and adds it to a set called $processed$. The \emph{graphC} algorithm ignores any connected components in the $processed$ set going forward. Next, \emph{graphC} restores the original signs of the target connected component before invoking the graphBplus algorithm again. If the improvement of the split is greater than $epsilon$, the split is adopted, and \emph{graphC} proceeds to the next connected component. Algorithm~\ref{alg-community} outlines the steps of the proposed community discovery algorithm. The $t_l$ is an optional time limit parameter for the algorithm. If $t_l = -1$, the algorithm will run until it meets the $\epsilon$ stopping criteria. 

\begin{algorithm}[!h]
\SetAlFnt{\small\sf}
\SetAlgoLined
\caption{Choose component to split}\label{alg-community}
\label{alg-hararysplit}
\KwData{Signed graph $\Sigma_{ij}$, spanning tree sampling method $M$, $I$, $\alpha$, $\beta$, $\epsilon$, $\Gamma$, Time limit $t_l$}
\KwResult{Clustering label assignment set $\mathcal{C} = \{C_v\}, v \in V$}
label\_counter=$0$\;
Call Alg.~\ref{alg-label} with inputs
$G,label\_counter,\mathcal{C}$\;
set R = $\emptyset$, processed = $\emptyset$\;
\While{true}{
\If{$ t \geq t_l$}{
break\;
}
label\_selected=-1 \;
terminate=true \;
\For{label $l, l \in \mathcal{C}$}{
\If{$|CC_l| \leq {Gamma}$}{
continue\;
}
\If{$l$ is not in set processed}{
label\_selected=$l$~terminate = false\;
}
}
\If{terminate == true}{
break\;
}
$\mathcal{C}_t = \mathcal{C}$~
label\_counter\_temp = label\_counter \;
$G^{FC}=$ Call Alg.~\ref{alg-bestsplit} ($\Sigma_{ij}$, Set R,label\_selected, $\mathcal{C}$,$\alpha$, $\beta$, $I$)\;
\For{$ CC \in G^{FC}$ }{
\For{vertex $i \in CC$}{
$\mathcal{C_i}$=label\_counter\;
}
label\_counter++\;
}
Compute $\mathcal{U}_{t+1,G}$ \;
\If{$\mathcal{U}_{t,G}-\mathcal{U}_{t+1,G} \leq \epsilon $}{
$\mathcal{C} = \mathcal{C}_t$~ label\_counter = label\_counter\_temp\;
Append label\_selected to processed and exit loop\;
}
Set $\mathcal{U}_{t,G} = \mathcal{U}_{t+1,G}$ $R = \emptyset$ \;}
\end{algorithm}

\begin{table}[!h]
\centering
\caption{Ten state-of-the-art leading methods from the literature.\label{tab-baseline}}
\begin{tabular}{c|c}
\toprule
\textbf{SOTA} & \textbf{Description} \\ \midrule
\textbf{Laplacian\_none}~\cite{knyazev2017signed} & spectral clustering using the signed graph Laplacian\\ 
\textbf{Laplacian\_sym}~\cite{knyazev2017signed} & spectral clustering using the symmetric Laplacian.\\
\textbf{BNC\_none}~\cite{2012chiang} & balanced normalized cuts.\\
\textbf{BNC\_sym}~\cite{2012chiang} & symmetric balanced normalized cuts.\\
\textbf{SPONGE\_none}~\cite{pmlr-v89-cucuringu19a} & SPONGE implementation.\\
\textbf{SPONGE\_sym}~\cite{pmlr-v89-cucuringu19a} & symmetric SPONGE implementation.\\
\textbf{Hessian}~\cite{Saade2014SpectralCO} & clustering based on signed Bethe Hessian.\\
\textbf{A\_sym}~\cite{signet_repo} & spectral clustering using symmetric adjacency matrix.\\
\textbf{dns}~\cite{snsdns} & clusters the graph using eigenvectors of the bns Laplacian matrix.\\
\textbf{sns}~\cite{snsdns} & clusters the graph using eigenvectors of the sns Laplacian matrix.\\ \bottomrule
\end{tabular}
\end{table}

\begin{figure}[!h]
\centering
  \includegraphics[width=0.5\textwidth]{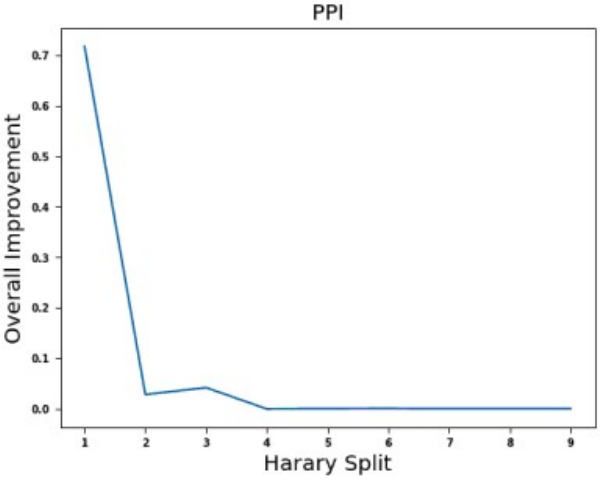}
  \caption{Example output of the overall improvements with each \textbf{committed} Harary split of every connected component of our proposed algorithm on the signed PPI graph.}
  \label{fig-exec}
\end{figure}

\textbf{Diminishing Returns (Fostering Scalability):} An intrinsic phenomenon in hierarchical clustering algorithms is that the deeper we go, the fewer potential improvements we get. Finding the best Harary cut of every connected component is computationally expensive, especially on massive signed graphs, as they might have millions of relatively small connected components, which could yield marginal improvements if any at all. This observation justifies using the optional $\Gamma$ parameter to save computation time. It represents a trade-off between efficiency and performance. If the connected component's size is less than $\Gamma$, the algorithm ignores it and will not find the best Harary cut even if it is not in the $processed$ set. Figure~\ref{fig-exec} illustrates a typical execution of the algorithm (on the signed PPI graph) and its gradually decreasing improvement over each Harary split. The overall improvement dropped from around 0.7 to a meager 0.0007 in the $5^{th}$ Harary cut. 

\begin{figure}[!h]
\centering
  \includegraphics[width=3in]{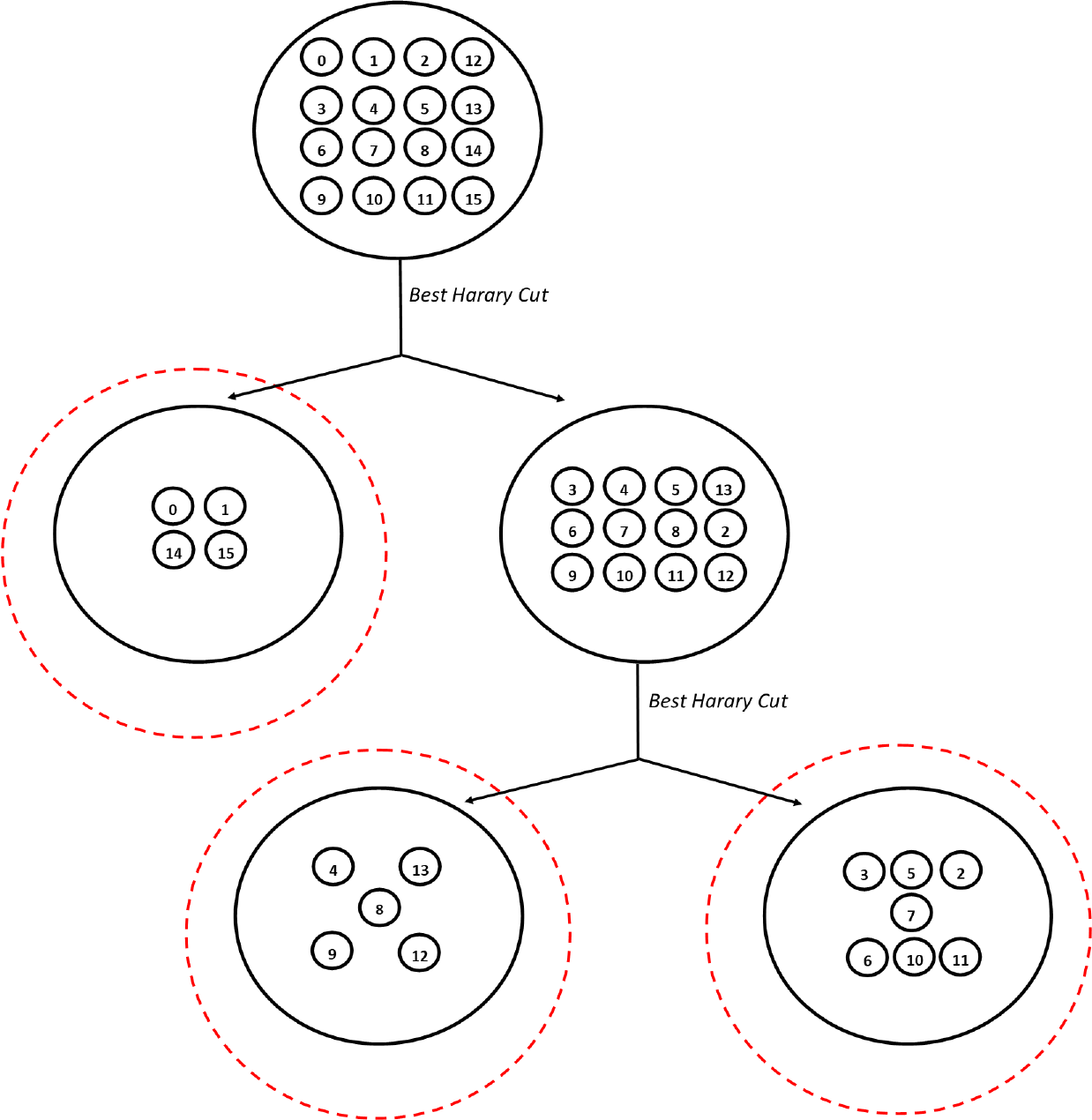}
  \caption{Execution of the \emph{graphC} algorithm on the Highland signed graph comprising 16 vertices and 58 edges.}
  \label{fig-Highland}
\end{figure}

\textbf{Illustrative Execution on Highland Tribes Graph:}
Figure~\ref{fig-Highland} shows the execution of our proposed algorithm. Assume that $I=$ 1000, $\alpha=$ 0.5 and $\beta=$ 1, $Gamma=$ 2, $\epsilon=$ 0.00000001, and $t_l=$ -1. First, the algorithm checks if it exceeded the time limit $t_l$; it proceeds since there is no time limit. Next, the \emph{graphC} algorithm checks whether the initial component at level 0 is in $processed$. The first pass directly proceeds to the next step because the $processed$ set is empty. Then, \emph{graphC} restores the original signs and finds the best Harary cut for the processed component. In the first iteration, that is the largest connected component of the graph, and the result is two connected components. This step is iteratively repeated for both components. After performing the Harary split, the algorithm stops splitting the current component when the improvement is less than $epsilon$. Recall that it undoes the last split and commits the connected components found so far, as illustrated in Figure~\ref{fig-Highland} with red dashed circles. The vertices within that component are given a unique final label. Eventually, the two components at level 1 also trigger the same stopping criteria, resulting in 3 clusters.

\section{Proof of Concept Implementation}

\subsection{Implementation and Setup}

 The implementations of the state-of-the-art methods listed in Table~\ref{tab-baseline} are in Python from the SIGNET repository \cite{signet_repo}. The parameters we used for all Konect signed graphs are $I =$ 1000, $\alpha =$ 0.5, $\beta =$ 1, $\epsilon =$ 0.00000001, $Gamma =$ 2, and $t_l =$ 100000 except for WikiConflict and WikiPolitics, where we chose $Gamma =$ 10 to speed up the execution. The parameters used for all Amazon signed graphs are $I =$ 50, $\alpha =$ 0.5, $\beta =$ 1, $\epsilon =$ 0.00000001, $Gamma =$ 40, and $t_l =$ -1. Next, we run the k-means++ clustering ten times with different centroid seeds for all spectral methods to cluster the eigenvectors. For all graphs, the number of communities $k$ was chosen based on two recent papers~\cite{he2022sssnet, 2022Survey}, as outlined in Table~\ref{tab-konectData}. If the signed graph does not have a known $k$ in the literature, we assign a $k$ value equal to the $k$ value of the most similar graph in size. We implemented the \emph{graphC} algorithm in C++ to automatically discover optimal communities without a predefined number of clusters $k$. The  \emph{graphC} finds and commits the optimal Harary cut for each connected component if it satisfies the following conditions: it is not in in the $processed$ set, does not exceed the time limit $t_l$, its size is more significant than $\Gamma$, and the Harary cut leads to an improvement greater than $\epsilon$. The code for running the leading ten methods in Table~\ref{tab-baseline} and for \emph{graphC} implementation is available at \url{https://anonymous.4open.science/r/graphC-2046/}. 
 
The operating system used for the experiments is Linux Ubuntu 20.04.3. The CPU is an 11th-generation Intel(R) Core(TM) i9-11900K @ 3.50GHz with 16 physical cores. It has one socket, two threads per core, and eight cores per socket. The architecture is x86\_x64. The GPU is an NVIDIA GeForce RTX 3070 with 8GB of memory. The driver version is 495.29.05, and the CUDA version is 11.5. 

\subsection{Signed Graph Datasets}

All signed graphs are pre-processed to purge self-edges, inconsistent edges, and duplicate edges (keeping only the first) and treat neutral edges as positive. Table~\ref{tab-konectData} describes the graphs derived as the Konect \cite{konect} benchmark as follows: {\em Highland} is the Konect signed social network of tribes of the Gahuku\-Gama alliance structure of the Eastern Central Highlands of New Guinea, from Kenneth Read \cite{1954Read}. {\em Congress} is a signed network where vertices are politicians speaking in the United States Congress, and a directed edge denotes that a speaker mentions another speaker. In the \emph{Chess} network, each vertex is a chess player, and a directed edge represents a game with the white player having an outgoing edge and the black player having an incoming edge, where the weight of the edge represents the outcome. \emph{BitcoinAlpha} is a user-user trust/distrust network from the Bitcoin Alpha platform for trading bitcoins, and \emph{BitcoinOTC} is a user-user trust/distrust network from the Bitcoin OTC platform for trading Bitcoins. \emph{WikiElec} is the network of users from the English Wikipedia that voted for and against each other in admin elections. The \emph{SlashdotZoo} graph is the reply network of the technology website Slashdot, where the vertices are users, and the edges are replies. The edges of \emph{WikiConflict} represent positive and negative conflicts between users of the English Wikipedia. emph{WikiPolitics} is an undirected signed network that contains interactions between the users of the English Wikipedia that have edited pages about politics. Each interaction, such as text editing and votes, is valued positively or negatively. \emph{Epinions} is the trust and distrust network of Epinions, an online product rating site. It incorporates individual users connected by directed trust and distrust links. Table~\ref{tab-konectData} lists the characteristics of the Konect signed graphs, as well as the following signed graph networks: (1) the \emph{TwitterRef} captures data from Twitter concerning the 2016 Italian Referendum. Different stances between users signify a negative tie, while the same stances indicate a positive link~\cite{Lai2018}; (2) the {\em Sampson25} models the sentiment over time between novice monks in a New England monastery captured by Sampson~\cite{1968Sampson}; and (3) the \emph{PPI} models the protein-protein interaction network~\cite{he2022sssnet}. \emph{WikiRfa} describes voting information for electing Wikipedia managers~\cite{he2022sssnet}.

Table~\ref{tab-amazonData} outlines the characteristics of the most significant connected components for the signed graphs derived from the Amazon ratings and reviews benchmark dataset. The Amazon dataset consists of \emph{seventeen} signed graphs derived from the Amazon rating and review files~\cite{2016Amazon2}. The dataset contains Amazon's product reviews and metadata from May 1996 and July 2014. The rating score of 1 to 5 is mapped to an edge between the user and the product as follows $(5,4) \rightarrow m^+$, $3 \rightarrow m$ (no sign), and $(2, 1) \rightarrow m^-$~\cite{2016Amazon2}. Table~\ref{tab-amazonData} outlines the characteristics of the most significant connected component. For example, clustering a bipartite graph of users and items in a recommendation system context can reveal groups of users with similar preferences and groups of items that are popular among certain users. Another example is clustering a bipartite graph of authors and papers, which can identify research communities and topics.

\section{Konect Signed Graphs Experiment}

\begin{table}[!ht]
\centering %\footnotesize
\caption{Konect + benchmark: Konect~\cite{konect} as well as TwitterReferendum~\cite{Lai2018}, PPI~\cite{he2022sssnet}, and WikiRfa~\cite{he2022sssnet} signed graphs. LCC stands for the largest connected component, $k$ is the predefined number of clusters for the spectral clustering methods, and the \emph{graphC} results list the number of clusters with at least five elements and the number of clusters with fewer than five elements. The last column is the number of splits produced after the algorithm's execution.}
\label{tab-konectData}
\begin{tabular}{l||r|r||r||r|r|r}
 & \multicolumn{2}{c||} {\bf LCC} & \multicolumn{1}{c||} {\bf $k$} & \multicolumn{3}{c}{GraphC} \\ 
\bf Dataset &\# \bf vertices&\# \bf edges&\# \bf clusters & \# \bf clusters$\geq$5 & \# \bf clusters$<$5&\# \bf splits \\ \hline \hline
\emph{Highland}& 16 & 58 & 3 & 1 & 2& 2\\ \hline
\emph{Sampson25}& 25 & 165 & 4 & 2 & 2& 3\\ \hline
\emph{Congress}& 219 & 521 & 11 & 2 &9 & 3\\ \hline
\emph{PPI}& 3058 & 11l860 & 10 & 29 & 887 &16\\ \hline
\emph{BitcoinAlpha} & 3l775 &14,120 & 10 & 14 & 201 &17\\ \hline
\emph{BitcoinOTC}& 5,875 & 21,489& 20 & 22 & 518 &10\\ \hline
\emph{Chess}& 7,115 & 55779& 30 & 46 & 1,014&41\\ \hline
\emph{TwitterRef}& 10l864& 251,396& 50 & 4 &234 &8 \\ \hline
\emph{SlashdotZoo}& 79l116 &467l731 & 100 & 245 &14,603 &53\\ \hline
\emph{Epinions}& 119l130 & 704,267& 100 & 584 &27,498 &237\\ \hline
\emph{WikiRfa}&7,634& 175787 &30 & 25 & 1,419&24\\ \hline
\emph{WikiElec}&7,066& 100l667& 30 & 38 & 1,528 &25\\ \hline
\emph{WikiConflict}& 113,123 &2,025,910 & 100 & 208 &66,083 &32\\ \hline
\emph{WikiPolitics}& 137,740 & 715,334 & 100 & 875 & 18,964&89 \\ 
\end{tabular}
\end{table}

%\textbf{SlashdotZoo} & {\ul \textbf{69}} & {\ul \textbf{86}} & 716 & \\
%\textbf{Epinions} & {\ul \textbf{80}} & {\ul \textbf{88}} & 2094 & \\
%\textbf{WikiConflict} & {\ul \textbf{94}} & {\ul \textbf{86}} & 1235 & & & & & & & & & & & & & & & & & & & & & & & & & & & & & & \\\hline
%\textbf{WikiPolitics} & {\ul \textbf{82}} & {\ul \textbf{85}} & 1279 & & & & & & & & & & & & & & & & & & & & & & & & & & & & & & \\

\begin{table*}[!ht]
\setlength\tabcolsep{2pt}
\centering %\footnotesize
\caption{Evaluation results of our proposed algorithm against ten different leading methods on 14 datasets in terms of ${pos}_{in}(G)$ and ${neg}_{out}(G)$, and time in seconds. Empty cells indicate the method failed to run on the dataset.}
\label{tab-results}
%\resizebox{2.05\columnwidth}{!}{
\begin{tabular}{p{2cm}||ccc|ccc|ccc|ccc|ccc|ccc}
\textbf{Method $\rightarrow$} & \multicolumn{3}{c|}{\emph{GraphC}} & \multicolumn{3}{c|}{\textbf{$\mathbf{A_{sym}}$}} & \multicolumn{3}{c|}{\textbf{$\mathbf{L_{none}}$}} & \multicolumn{3}{c|}{\textbf{$\mathbf{L_{sym}}$}} & \multicolumn{3}{c|}{\textbf{$\mathbf{BNC_{none}}$}} & \multicolumn{3}{c}{\textbf{$\mathbf{BNC_{sym}}$}} \\ 
Dataset $\downarrow$ & \textbf{pos} & \textbf{neg} & \textbf{t} & \textbf{pos} & \textbf{neg} & \textbf{t} & \textbf{pos} & \textbf{neg} & \textbf{t} & \textbf{pos} & \textbf{neg} & \textbf{t} & \textbf{pos} & \textbf{neg} & \textbf{t} & \textbf{pos} & \textbf{neg} & \textbf{t} \\ \midrule
\textbf{Highland} & {\ul \textbf{93}} & {\ul \textbf{100}} & 0.1 & {\ul 93} & {\ul 100} & 0.1 & {\ul 93} & {\ul 100} & 0.1 & {\ul 93} & {\ul 100} & 0.1 & {\ul 93} & {\ul 100} & 0.1 & {\ul 93} & {\ul 100} & 0.1 \\
\textbf{Sampson25} & {\ul \textbf{70}} & {\ul \textbf{90}} & 0.2 & 60 & 91 & 0.04 & 59 & 79 & 0.1 &63 & 92 & 0.03 & 78 & 65 & 0.03 & 62 & 91 & 0.04 \\
\textbf{Congress} & {\ul \textbf{93}} & {\ul \textbf{100}} & 0.77 & 70 & 100 & 0.11 & 86 & 97 & 0.12 & 24 & 100 & 0.11 & 39 & 77 & 0.13 & 93 & 97 & 0.14\\ 
\textbf{PPI} & {\ul \textbf{81}} & {\ul \textbf{98}} & 9 & 100 & 2 &  1 & 100 & 0.98 & 1 & 93 & 10 & 1 & 100 & 0.98 & 1 \\
\textbf{BitcoinAlpha} & {\ul \textbf{81}} & {\ul \textbf{86}} & 14 & 100 & 0.5 & 2 & 30 & 68 & 3 & 72 & 24 & 3 & 55 & 66 & 2 & 100 & 3 & 2\\
\textbf{BitcoinOTC} & {\ul \textbf{82}} & {\ul \textbf{90}} & 22 & 71 & 42 & 4 & 21 & 80 & 7 & 61 & 43 & 4 & 86 & 61 & 4 & 100 & 35 & 4 \\ 
\textbf{Chess} & {\ul \textbf{39}} & {\ul \textbf{84}} & 45 & 100 & 0.17 & 7 & & & & & & & 65 & 34 & 7 & 100 & 0.21 & 7\\ 
\textbf{TwitterRef} & {\ul \textbf{89}} & {\ul \textbf{100}} & 446 & 65 & 90 & 31 & & & & 67 & 94 & 31 & 23 & 93 & 30 & 100 & 0.66 & 31 \\ 
\textbf{WikiRfa} & {\ul \textbf{61}} & {\ul \textbf{81}} & 83 & 79 & 30 & 18 & 100 & 0.003 & 19 & 74 & 35 & 18 & 100 & 2 & 18 & 100 & 0.01 & 18\\ 
\textbf{WikiElec} & {\ul \textbf{70}} & {\ul \textbf{79}} & 60 & 19 & 65 & 11 & 100 & 0.05 & 21 & 100 & 0.05 & 11 & 51 & 58 & 11 & 1 & 0.11 & 11\\ \bottomrule
\end{tabular} 
\begin{tabular}{p{2cm}||ccc|ccc|ccc|ccc|ccc|ccc} \toprule
\textbf{Method $\rightarrow$} & \multicolumn{3}{c|}{\emph{GraphC}} &\multicolumn{3}{c|}{\textbf{Hessian}}  & \multicolumn{3}{c|}{\textbf{$\mathbf{SPO_{none}}$}} & \multicolumn{3}{c|}{\textbf{$\mathbf{SPO_{sym}}$}} & \multicolumn{3}{c|}{\textbf{dns}} &\multicolumn{3}{c}{\textbf{sns}} \\ 
Dataset $\downarrow$ & \textbf{pos} & \textbf{neg} & \textbf{t} & \textbf{pos} & \textbf{neg} & \textbf{t} & \textbf{pos} & \textbf{neg} & \textbf{t} & \textbf{pos} & \textbf{neg} & \textbf{t} & \textbf{pos} & \textbf{neg} & \textbf{t} & \textbf{pos} & \textbf{neg} & \textbf{t} \\ \midrule
 
\textbf{Highland} & {\ul \textbf{93}} & {\ul \textbf{100}} & 0.1 & {\ul 93} & {\ul 100} & 0.2 & {\ul 93} & {\ul 100} & 0.19 & {\ul 93} & {\ul 100} & 0.17 & {\ul 93} & {\ul 100} & 0.2 & {\ul 93} & {\ul 100} & 0.1 \\ 
\textbf{Sampson25} & {\ul \textbf{70}} & {\ul \textbf{90}} & 0.2 & 63 & 93 & 0.04 & 63 & 92 & 0.04 & 64 & 92 & 0.04 & 70 & 85 & 0.03 & 70 & 82 & 0.03 \\
\textbf{Congress} & {\ul \textbf{93}} & {\ul \textbf{100}} & 0.77 & 62 & 100 & 0.11 & 70 & 84 & 0.14 & 67 & 100 & 0.12 & 94 & 95 & 0.12 & 98 & 33 & 0.12 \\ 
\textbf{PPI} & {\ul \textbf{81}} & {\ul \textbf{98}} & 9 & 79 & 35 & 1 & 100 & 0.1 & 1 & 100 & 1 & 1 & 100 & 0.72 & 1 & 100 & 0.77 & 1 \\
\textbf{BitcoinAlpha} & {\ul \textbf{81}} & {\ul \textbf{86}} & 14 & 37 & 89 & 2 & 91 & 3 & 3
& 100 & 5 & 3  & 100 & 4 & 2  & 100 & 7 & 2 \\
\textbf{BitcoinOTC} & {\ul \textbf{82}} & {\ul \textbf{90}} & 22 & 30 & 94 & 4 & 62 & 30 & 8 & 100 & 28 & 4 & 100 & 8 & 4 & 100 & 8 & 4 \\
\textbf{Chess} & {\ul \textbf{39}} & {\ul \textbf{84}} & 45 & 40 & 70 & 7 & 100 & 0 & 12 & 100 & 0.51 & 8 & 100 & 0.2 & 7 & 100 & 0.2 & 7 \\ 
\textbf{TwitterRef} & {\ul \textbf{89}} & {\ul \textbf{100}} & 446 & 11 & 98 & 31 & 97 & 4 & 60 & 65 & 94 & 31 & 100 & 0.6 & 30 & 100 & 0.55 & 31 \\
\textbf{WikiRfa} & {\ul \textbf{61}} & {\ul \textbf{81}} & 83 & 54 & 59 & 18 &  9 & 11 & 2 & 91 & 6 & 19 & 100 & 0.02 & 18 & 100 & 0.01 & 18 \\ 
\textbf{WikiElec} & {\ul \textbf{70}} & {\ul \textbf{79}} & 60 & 19 & 92 & 11  & 92 & 12 & 20 & 1 & 0.74 & 11 & 1 & 0.17 & 11 & 1 & 0.16 & 11 \\ \bottomrule
\end{tabular}
\end{table*}

\begin{figure}[!h]
  \centering
  \includegraphics[scale=0.65]{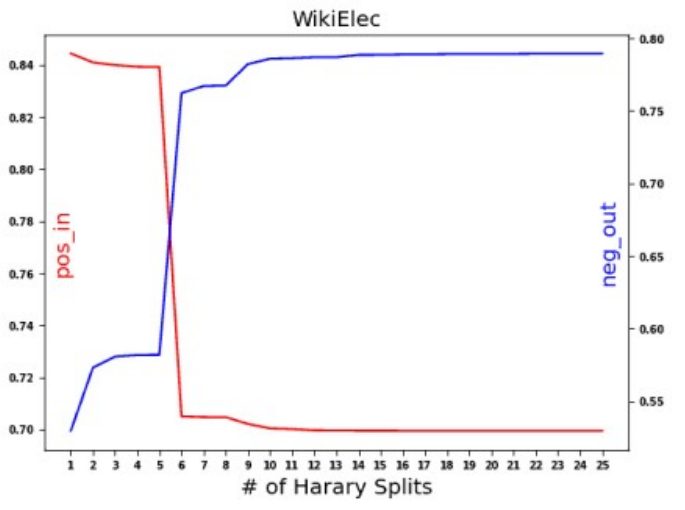}
  \caption{${pos}_{in}$ and ${neg}_{out}$ for WikiElec with each Harary split during algorithm execution}
   \label{fig-wikielections}
\end{figure}

The \emph{graphC} algorithm also counts the number of clusters (including isolated vertices) and the number of Harary split operations executed for various signed networks. We analyze and visualize the impact of increasing the number of subsequent Harary splits on the overall improvement in Equation~\ref{eq-overall_loss} using the two evaluation metrics $pos_{in}$ (Eq.~\ref{eq-PosIn}) and $neg_{out}$ (Eq.~\ref{eq-NegOut}). Figure~\ref{fig-wikielections} shows the results of running our algorithm on the WikiElec signed graph. Initially, $pos_{in} = 1.0$ and  $neg_{out} = 0.0$ as all edges belong to a single cluster. Our algorithm divides the connected component for each split to minimize the loss of $pos_{within}$ and maximize the gain of $neg_{between}$. Figure~\ref{fig-wikielections} also shows that reducing $pos_{within}$ with each split is inevitable, but the rate at which $neg_{between}$ increases is much higher than the rate at which $pos_{within}$ decreases during the first few splits, and the overall measure of the quality of the clusters improves. Afterward, the performance starts to plateau, with diminishing gains as the algorithm makes deeper cuts into the connected components. The parameter $\Gamma$ allows \emph{graphC} to stop before reaching the plateau and saves computation time.

The methods with predefined $k$ parameters force isolated vertices into arbitrary clusters to satisfy the $k$-value constraint~\cite{2022Cluster}. In contrast, \emph{graphC} allows for the isolated vertices to be automatically detected as long as the improvement in Equation~\ref{eq-overall_loss} exceeds the threshold $\epsilon$ that specified that a vertex may split from its connected component if the overall improvement increases by $\epsilon$ value. The \emph{graphC} algorithm finds a clustering that consistently achieves a good compromise between ${pos}_{in}(G)$ and ${neg}_{out}(G)$ (pos and neg in this table are percentages) on the Konect dataset with \textbf{both} having high values unlike other methods where they fail or solely focus on either evaluation metric. Table~\ref{tab-konectData} summarizes the number of clusters and the number of Harary split operations performed on the Konect signed graphs, where the best results are underlined. 

Next, we evaluate the performance and execution time of our algorithm against ten leading methods from the literature, which is also shown in Table~\ref{tab-results}. Empty cells in the table indicate two possible issues: either the state-of-the-art algorithm took more than two days to run or failed due to non-convergence of the eigenvectors during the spectral decomposition, particularly for large signed graphs. In all tested signed graphs, regardless of size or density, our algorithm outperforms the state-of-the-art in terms of \( pos_{in} \) and \( neg_{out} \). The execution time of our algorithm is longer than some other methods on certain graphs (e.g., WikiElec) because the $ k$ value in these methods is relatively small, leading to faster execution times. However, the state-of-the-art methods suffer significant performance degradation as the graph size and $k$-value increase while our algorithm maintains a more consistent execution time.

\section{Amazon Ratings and Reviews Modeling}

We map the Amazon ratings and reviews~\cite{2016Amazon2} to signed graphs with negative edges for ratings of 0, 1, and 2, no sign if the rating is 3, and a positive sign for ratings of 4 and 5. We demonstrate that our algorithm can scale to a signed network with millions of vertices and edges by running it on the Amazon signed graphs listed in Table~\ref{tab-amazonData}. For instance, \emph{graphC} successfully runs the largest Amazon graph (Books) in 31 hours while yielding high $pos_{in}$ and $neg_{out}$ percentages. %Therefore, our algorithm can detect clusters automatically in a scalable fashion.
As Table~\ref{tab-amazonData} shows, the bipartite graph of Amazon reviews/ratings has a skewed distribution of vertices and edges, where most of the vertices are sparsely connected, and only a few vertices dominate the market. For instance, in the Books signed graph, about 1 million clusters contain less than five users/books, and only 32 thousand clusters have more than five users/books. Although these clusters contain a mix of users and items, %and no quantification of the number of users each item has in its cluster is done,
this may correlate with the ``Long Tail Phenomenon''~\cite{anderson2006long}, where a large number of products or services that are not very popular can jointly have a significant share of the market that may overtake or be comparable to the current bestsellers. 

\begin{table*}[!ht]
\setlength\tabcolsep{2pt}
\centering %\footnotesize
    \caption{Amazon ratings and reviews~\cite{2016Amazon2} mapped to signed graphs. The \emph{graphC} results include ${pos}_{in}(G)$ and ${neg}_{out}(G)$ (as percentages) and the resulting number of clusters with at least five elements as well as the number of clusters with less than five elements.}
  \label{tab-amazonData}
  \begin{tabular}{l||r|r|r|r|r|r|r|r}
     \bf Amazon & \multicolumn{2}{c|} {\bf LCC} & \multicolumn{6}{c} {\bf GraphC}\\ \hline \hline
    \bf Ratings & \bf \# vertices& \bf \# edges &\textbf{pos}& \textbf{neg}&\textbf{\# splits}& \textbf{time (s)} & \# \bf clusters$\geq$5 & \# \bf clusters$<$5 \\ \hline
    Books & 9,973,735 & 22,268,630 & 73 & 86 & 22 &115,561 & 31,988& 1,003,734 \\ \hline
    Electronics & 4,523,296 & 7,734,582 &88 & 80 & 7 & 23,516 & 9,208 & 862,970 \\ \hline
    Jewelry & 3,796,967 & 5,484,633 & 81 & 90 &11 & 25,650 & 15,802 & 648,695 \\ \hline
    TV & 2,236,744 & 4,573,784 & 74 & 87 &17 &19,019 & 4,456 & 281,653 \\ \hline
    Vinyl & 1,959,693 & 3,684,143 & 74 & 87  & 16& 16,420 & 5,718 & 169,493 \\ \hline
    Outdoors & 2,147,848 & 3,075,419& 92 &80  &15 &13,613 &12,294& 393,579 \\ \hline
    AndrApp & 1,373,018 & 2,631,009 &77 & 89 & 24 & 1,642 & 1,462 & 205,336 \\ \hline
    Games & 1,489,764 & 2,142,593 & 82 & 22 &1,662 & 280,356& 8,111& 272,245\\ \hline
    Automoto & 950,831 & 1,239,450&94&79 & 13 & 783 & 8,515&202,749 \\ \hline
    Garden& 735,815 & 939,679 & 93&85& 13  & 436 &4,636 & 153,452 \\ \hline
    Baby & 559,040 & 892,231&79 & 91&14 & 489 & 2,367 & 94,474 \\ \hline
    Music & 525,522 & 702,584& 87& 83 &23 &571& 6,124 & 109,287\\ \hline
    Video & 433,702 & 572,834 &85 &93& 14 & 364 &1,401 & 46,481 \\ \hline
    Instruments & 355,507 & 457,140 &90 &85 &15 & 281 & 3,536 & 61,964 \\ \hline 
    \hline
    \bf Reviews & \bf \# vertices& \bf \# edges &\textbf{pos}& \textbf{neg}&\textbf{\# splits}& \textbf{time (s)} & \bf clusters$\geq$5 & \# \bf clusters$<$5 \\ \hline
    Core Music & 9,109 & 64,706 &0.520 & 0.848  & 14 & 9.57& 38 & 746 \\ \hline
    Core Video& 6,815 & 37,126 &0.572 & 0.852 & 13 & 5.54& 26& 737 \\ \hline
    Core Instrum& 2,329 & 10,261& 0.510 & 0.928 & 8 &1.90& 20 & 159 \\ 
  \end{tabular}
\end{table*}

\subsection{Parameters Study \& Analysis}
\begin{figure*}[!ht]
  \centering
  \includegraphics[width=\textwidth]{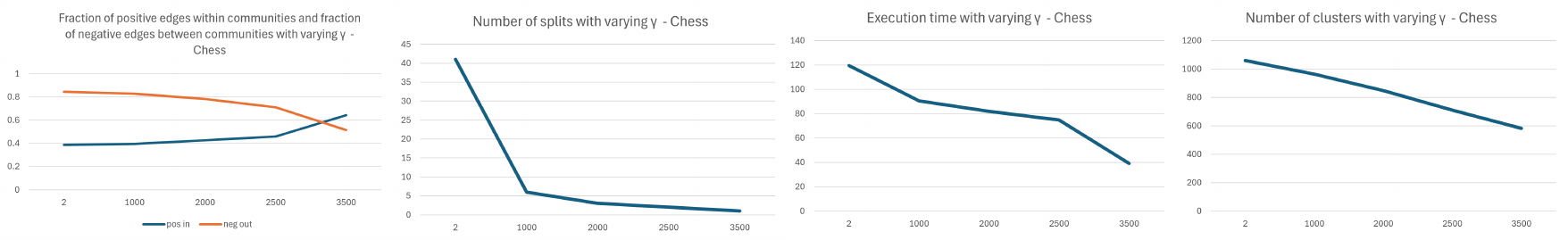}
  \caption{The effect of varying $\Gamma$ on the number of clusters, clustering quality, execution time, and Harary splits on the Chess signed graph. The y-axis is ${pos}_{in}$ and ${neg}_{out}$ as in equations~\ref{eq-PosIn} and~\ref{eq-NegOut}, respectively, the number of splits, the execution time, and the number of communities for the four panels (from left to right). The x-axis is the $\Gamma$ value. The \emph{graphC} algorithm ignores connected components of size less than $\Gamma$.}
  \label{fig-gamma}
\end{figure*}
\begin{figure*}[!ht]
  \centering
  \includegraphics[width=\textwidth]{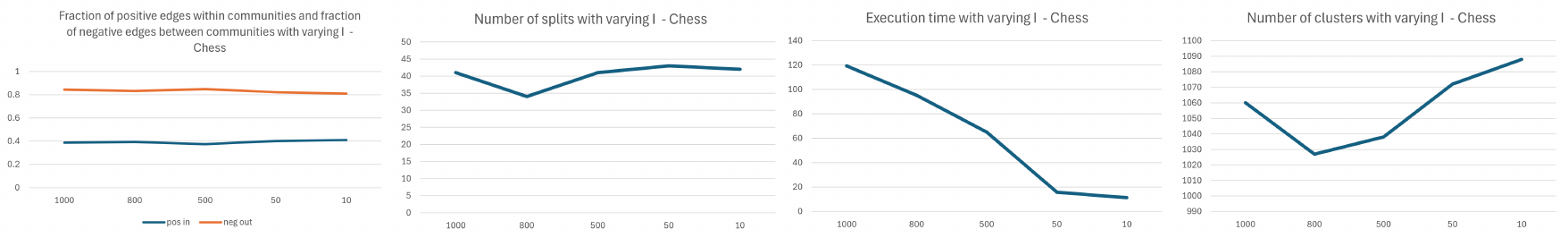}
  \caption{The effect of varying $I$ (number of iterations per connected component) on the number of clusters, clustering quality, execution time, and Harary splits on the Chess signed graph. The y-axis is ${pos}_{in}$ and ${neg}_{out}$ as in Equations~\ref{eq-PosIn} and \ref{eq-NegOut}, respectively, and the number of splits, the execution time, and the number of communities for the four panels (from left to right). The x-axis is the $I$ value.}
  \label{fig-iter}
\end{figure*}
\begin{figure*}[!ht]
  \centering
  \includegraphics[width=\textwidth]{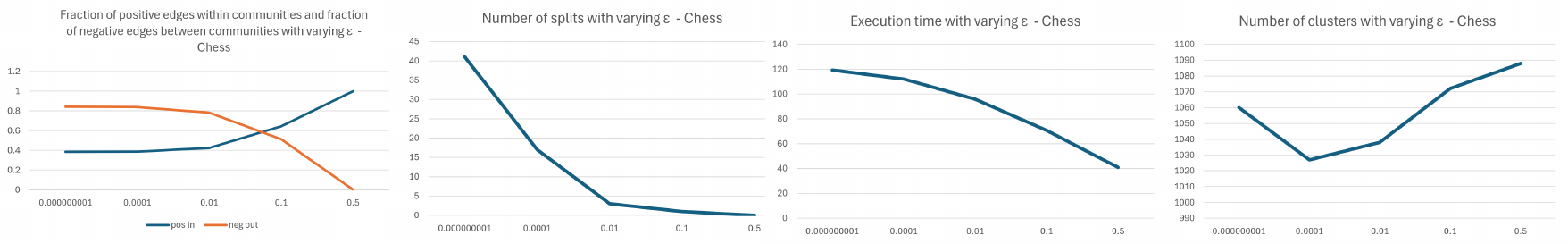}
  \caption{The threshold $\epsilon$ specifies the minimum improvement value for the set split. This image summarizes the effect of varying $\epsilon$ on the number of clusters, clustering quality, execution time, and Harary splits on the Chess signed graph. From left to right, the y-axis is ${pos}_{in}$ and ${neg}_{out}$ as in Equations~\ref{eq-PosIn} and \ref{eq-NegOut}, respectively.}
  \label{fig-epsilon}
\end{figure*}

To better understand the parameters $I$, $\Gamma$, and $\epsilon$ and their effect on the clustering process, we run \emph{graphC} on the Chess signed graph. First, we start by running the algorithm with $Gamma \in \{2, 1000, 2000, 2500, 3500\}$ while fixing the other parameters at $I=$ 50, $\alpha =$ 0.5, $\beta =$ 1, $\epsilon =$ 0.00000001, and $t_l =$ -1. As the four panels of Figure~\ref{fig-gamma} show, increasing $\Gamma$ deteriorates the clustering quality while the number of splits and the execution time decrease, resulting in a lower number of communities. A higher value of $\Gamma$ means that the algorithm will disregard even larger connected components that have the potential to improve the overall quality of the clustering if split. In return, we save computational time because the algorithm does not search for the best split for a connected component of size smaller than $\Gamma$.

Second, we run the algorithm with $I \in \{1000, 800, 500, 50,10 \}$ with the other parameters fixed at $Gamma=$ 2, $\alpha =$ 0.5, $\beta =$ 1, $\epsilon =$ 0.00000001, and $t_l =$ -1. According to the four panels of Figure~\ref{fig-iter}, decreasing the number of iterations when searching for the best Harary split for a connected component gradually degrades the overall quality of the clustering, reducing the execution time tremendously. In contrast, the number of splits and clusters seems to stay the same or fluctuate around 41 splits and 1060 clusters. Finally, we run our algorithm with $\epsilon \in \{0.000000001, 0.0001, 0.01, 0.1,0.5 \}$ with the other parameters fixed at $Gamma=$ 2, $\alpha =$ 0.5, $\beta =$ 1, $I =$ 1000, and $t_l =$ -1. The threshold $\epsilon$ specifies minimal overall improvement for the set split in the \emph{graphC} algorithm. Thus, increasing $\epsilon$ reduces the probability of the split, as there is less chance improvement will be higher than $\epsilon$, as illustrated in the four panels of Figure~\ref{fig-epsilon}. Thus, higher values of $\epsilon$ degrade the quality of the clustering process and further reduce the execution time of the algorithm. The number of clusters in this ablations study stays about the same ($1060$) for different numbers of iterations $I$. In summary, high $I$ and low $\Gamma$ and $\epsilon$ values produce the most optimal clustering for the given resources regardless of the signed graph's type, density, or size.

\section{Conclusion}

The structure of communities within a network can reveal hidden patterns and insights about the relationships and dynamics in complex systems. Community detection is widely used in various fields, from biology to social networks, to extract such information. Recent research has pointed out challenges in recovering community structures in sparse networks, especially when using spectral methods. Current issues with signed-graph clustering algorithms include selecting the right number of clusters $k$, eigenvalue contamination, lack of scalability to large graphs, and dependence on ground truth for clustering validation. In response to these issues, we propose \emph{graphC}, a new, scalable, hierarchical clustering algorithm for signed networks that automatically detects the optimal number of clusters without needing a predefined \(k\). % based on the \emph{graphBplus} algorithm.
Our approach is flexible and supports optional parameters to control outcomes such as the number of isolated vertices detected.

Additionally, we introduce the parameter \( \Gamma \) to balance efficiency and scalability. We compare our algorithm to the leading methods from the literature using the \( pos_{in} \) and \( neg_{out} \) metrics. The results show that our algorithm consistently performs well on all tested datasets, achieving higher values for both metrics than the state-of-the-art methods, which struggle with scaling and producing high-quality cluster assignments. Future work is to parallelize our algorithm to speed up its execution and to integrate consensus features proposed in previous work~\cite{2021Cloud} to enhance clustering quality.

\end{document}